\documentclass[aps,preprint,superscriptaddress]{revtex4-1}

\usepackage[final]{pdfpages}
\usepackage{amsmath,amsthm,amsfonts,amscd} 
\usepackage{graphicx}
\usepackage{hyperref}	% hyperref allows citing websites via url
\usepackage{color}		% Allows \textcolor{red}{Hello world}
\usepackage[normalem]{ulem}	% Allows strikethrough \sout{text goes here}
\usepackage{mathtools}  %Allows some summation features used here.
\usepackage{braket}	%Allows Dirac notion

\begin{document}

\title{Simultaneous, full characterization of a single-photon state using semiconductor quantum-dot light}

%\author{Tim Thomay$^{1,*}$, Sergey V. Polyakov$^2$, Olivier Gazzano$^{1}$, Elizabeth Goldschmidt$^{1,2,\dagger}$, Vivien Loo$^{1,\ddagger}$ \& Glenn S. Solomon$^{1,2}$}

\author{Tim Thomay}
\affiliation{Joint Quantum Institute, National Institute of Standards and Technology, \& University of Maryland, Gaithersburg, MD, USA.}
%\altaffiliation{current address: Department of Electrical Engineering, University at Buffalo, State University of New York, Buffalo, NY, USA}
\author{Sergey V. Polyakov}
\affiliation{National Institute of Standards and Technology, Gaithersburg, MD, USA.}
\author{Olivier Gazzano}
\affiliation{Joint Quantum Institute, National Institute of Standards and Technology, \& University of Maryland, Gaithersburg, MD, USA.}
\author{Elizabeth Goldschmidt}
%\altaffiliation{current address: US Army Research Laboratory, Adelphi, MD, USA.}
\affiliation{Joint Quantum Institute, National Institute of Standards and Technology, \& University of Maryland, Gaithersburg, MD, USA.}
\author{Vivien Loo}
\affiliation{Joint Quantum Institute, National Institute of Standards and Technology, \& University of Maryland, Gaithersburg, MD, USA.}
%\altaffiliation{current address:  Laboratoire Kastler Brossel, UPMC-Sorbonne Universit\'es, CNRS, ENS-PSL Research University, Coll\`ege de France, Paris, Franc}
\author{Glenn S. Solomon}
\email{Corresponding author: gsolomon@umd.edu}
\affiliation{Joint Quantum Institute, National Institute of Standards and Technology, \& University of Maryland, Gaithersburg, MD, USA.}
\affiliation{National Institute of Standards and Technology, Gaithersburg, MD, USA.}

%\date{\today}
 
\begin{abstract}
As single-photon sources become more mature and are used more often in quantum information, communications and measurement applications, their characterization becomes more important. 
Single-photon-like light is often characterized by its brightness, and two quantum properties: the single-photon composition  and the photon indistinguishability. While it is desirable to obtain these quantities from a single measurement, currently two or more measurements are required. 
Here, we simultaneously determine the brightness, the single photon purity, the indistinguishability, and the statistical distribution of Fock states to third order for a quantum light source. 
The measurement uses a pair of two-photon ($n=2$) number-resolving detectors. $n>2$ number-resolving detectors provide no additional advantage in the single-photon  characterization. 
The new method extracts more information per experimental trial than a conventional measurement for all input states, and is particularly more efficient for statistical mixtures of photon states. 
%Finally, the platform can be used as a metrology testbed, or as an elementary block of a linear optical quantum protocol, and as shown here can be evaluated using a boson sampling method. 
Thus, using this n=2, number-resolving detector scheme will provide advantages in a variety of quantum optics measurements and systems.
\end{abstract}

\maketitle	


%\thanks{}


%\date{\today}

%\pacs{xxx,yyy,zzz}
%\keywords{}

%\maketitle

%\section{}
%\subsection{}
%\subsubsection{}

\section*{Introduction} 
\vspace*{-1em}
Single-photon light is a central element of emerging quantum information systems such as quantum repeaters~\cite{sangouard2011,muralidharan2015} and bosonic logic~\cite{broome2013,spring2013,tillmann2013,crespi2013}. This nonclassical light is also used in quantum measurement protocols.
Such protocols offer advantages over classical measurement protocols for classical properties~\cite{franson_pra1992,steinberg_prl1992}, such as in accuracy and sensitivity~\cite{caves1980,dolinar_mit1973,cook_nature2007} and, clearly quantum measurement protocols are essential to access quantum properties.

Light has been traditionally characterized by its coherence properties through a series of normalized Glauber functions $g^{(n)}$, where $2n$ is the field correlation order \cite{glauber1963}. 
%In situations where intensity is measured and normalization is important, these are recast as $g^{(n)}$, where $n$ is the intensity correlation order.
The brightness is given by the unnormalized $g^{(1)}$ function (historically denoted $G^{(1)}$ \cite{glauber1963}) and the normalized second-order correlation function $g^{(2)}$ gives the likelihood of two-photon correlations.  
%For ideally, pure single-photon states $g^{(2)} \equiv 0$. 
When two photon correlations are nonzero ($g^{(2)} \neq 0$), as is often the case, it is necessary to evaluate higher order correlations~\cite{zhang2012,goldschmidt_pra2013}.
In general, to measure the photon-state statistics to $n^{th}$ order, normalized $n^{th}$-order correlations could be measured using  a single, appropriately fast $n^{th}$-order number resolving detector, if such a detector were available (see Fig. 1a)~\cite{tes,Lita2008,Levine2012}.
Alternatively, $n$ single-photon detectors can be used with beam-splitters in place of an $n$ number resolving detector~\cite{stevens2014third}.
For example, a $n=2$ number-resolving detector can be replaced by two single-photon detectors and a beam-splitter (Fig. 1b), and photon detections between the two detectors can be correlated \cite{HBT_nature1956}, as discussed in the next section.
\begin{figure}[t!]
       \begin{minipage}[ht!]{1\linewidth}
	\begin{center}
	\vspace*{1em}
	\includegraphics[width=4.0in]{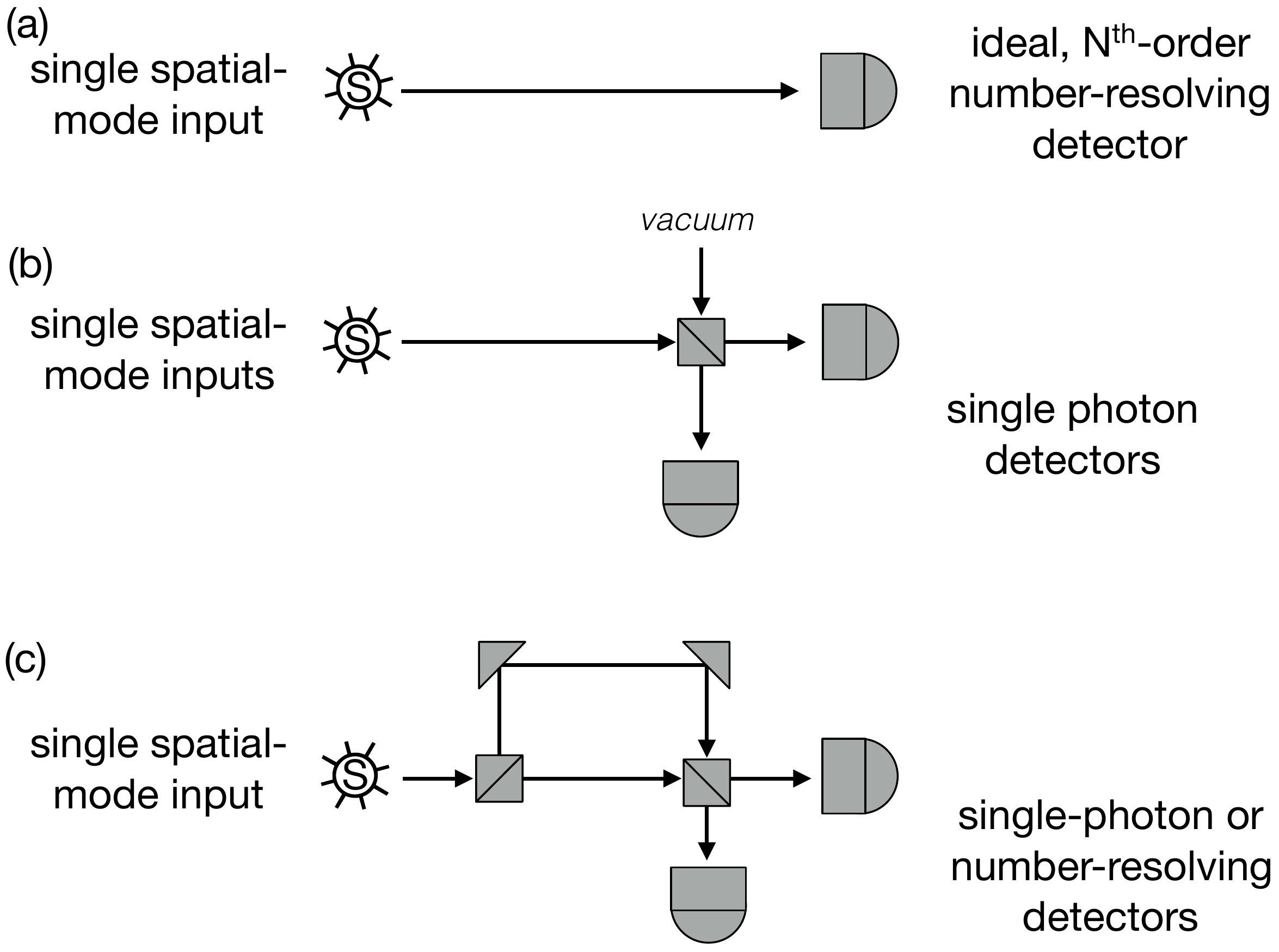}
	\end{center}
	\end{minipage}
	\vspace*{0.5em}
    \caption{Second-order characterization of light with single, spatial-mode inputs. (a)  The photon-state statistics can be measured to $n^{th}$ using  a single, appropriately fast $n^{th}$-order number resolving detector. Such a detector does not exist. (b) Instead of using a $n=2$ number-resolving detector, correlation measurements to $2^{nd}$ order can be made with two single-photon (not number resolving) detectors and a beam-splitter. The vertical beamsplitter input is vacuum.  (c) An unbalanced interferometer can be used to interfere replicas of an input state to measure the indistinguishability from second-order correlations. If single photon detectors are used an additional measurement; for instance, like the one in (b), must be made for normalization.}
\label{fig1new}
\end{figure}

In many quantum information applications; for instance, quantum repeater and bosonic sampling, the single photon state must also be indistinguishable. 
The indistinguishability is measured by interfering replicas of the photon state, sampled at different times or positions~\cite{hong_prl1987}, and can be measured with an unbalanced interferometer  (see Fig. 1c).
However, in an unbalanced interferometer scheme, the result must account for the single-photon nature of the light~\cite{flagg_prl2012}, requiring additional information.
Thus, to fully characterize the quantum state of the source--the photon number state and the indistinguishability--at least two distinct set-ups and measurements are required. Besides the obvious inefficiencies in changing set-ups and acquiring separate measurement results, the quantities defining the photon state are never evaluated together.

Here, we show a method to simultaneously access brightness, number-state statistics and indistinguishability. 
This ensures that these quantities are derived from the same measurement set, and that all aspects of the source and measurement are identical. 
A main feature of this simulantaneous second-order correlation (Sg2) approach is the use of $n=2$ number-resolving detectors in place of single-photon avalanche detectors (SPADs).
Number-resolving detectors have been demonstrated by several research groups~\cite{rosenberg2005,divochiy2008}, and they will likely be commercially available in the near future.
There are other important features of Sg2.
Using a pair of two-photon number resolving detectors, one can also simultaneously provide a complete characterization of the photon state to third order.
We also show that the photon-number resolved measurement intrinsically collects more information about an input photonic state than a similar measurement made with conventional detectors.
Thus, the Sg2 method is more efficient than a combination of conventional measurements, and becomes even more efficient as the single-photon purity degrades. 
Finally, when we substitute SPADs and beam-splitters for number-resolving detectors, the Sg2 layout mimics a simple linear optical circuit. 
The Sg2 scheme can be used to model such circuits or incorporated within them for local metrology testing.

\section*{Second-order correlations}
\vspace*{-1em}
The use of two detectors and a beam splitter to measure the second-order normalized correlation function goes back to measurements by Hanbury Brown and Twiss ~\cite{HBT_nature1956}, and we denoted it $g^{(2)}_{\rm HBT}$.
In the $g^{(2)}_{\rm HBT}$ measurement~\cite{HBT_nature1956}, a single  spatial mode is incident on one port of a beamsplitter and two detectors measure coincidences in the two output ports to assess if more than one photon is present. 
If $\tau$ is the difference in detection times for the two detectors, $g^{(2)}_{\rm HBT}[\tau=0] < 1$ is the hallmark of a quantum state. 
One value of $g^{(2)}_{\rm HBT}$, $g^{(2)}_{\rm HBT}[0] = 0$, represents a unique state. 
For this case, the non-vacuum component of the light is comprised of only single photons~\cite{mandel-wolf_1995,goldschmidt_pra2013}. 
$1 > g^{(2)}_{\rm HBT}[0] > 0$ signifies non-classical light with some multi-photon component, and the greater the value of $g^{(2)}_{\rm HBT}$ the higher the proportion of the multi-photon component in the source~\cite{kimble_prl1977}. 
Thus, for a source that is expected to provide single photons, $g^{(2)}_{\rm HBT}$ is used as a metric for single photon purity. 
This is not the purity of the $n=1$ number state, since the vacuum component is not represented in $g^{(2)}_{\rm HBT}$, but the purity against $n>1$ population.

\begin{figure}[h!]
       \begin{minipage}[c]{0.56\linewidth}
	\begin{center}
	\includegraphics[width=3.4in]{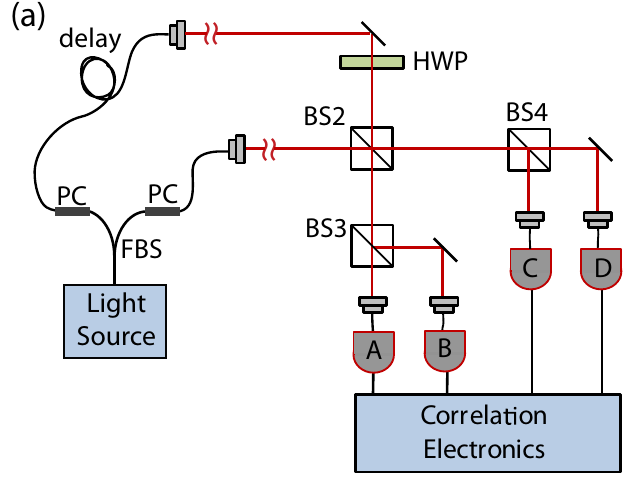}
	\end{center}
%	\vspace*{0.5em}
	\end{minipage}
       \begin{minipage}[h!]{0.39\linewidth}
	 \centering
	 	\begin{minipage}[h!]{1\linewidth}
		 \centering
	%	 \vspace*{-1em}
		\includegraphics[height=1.35in]{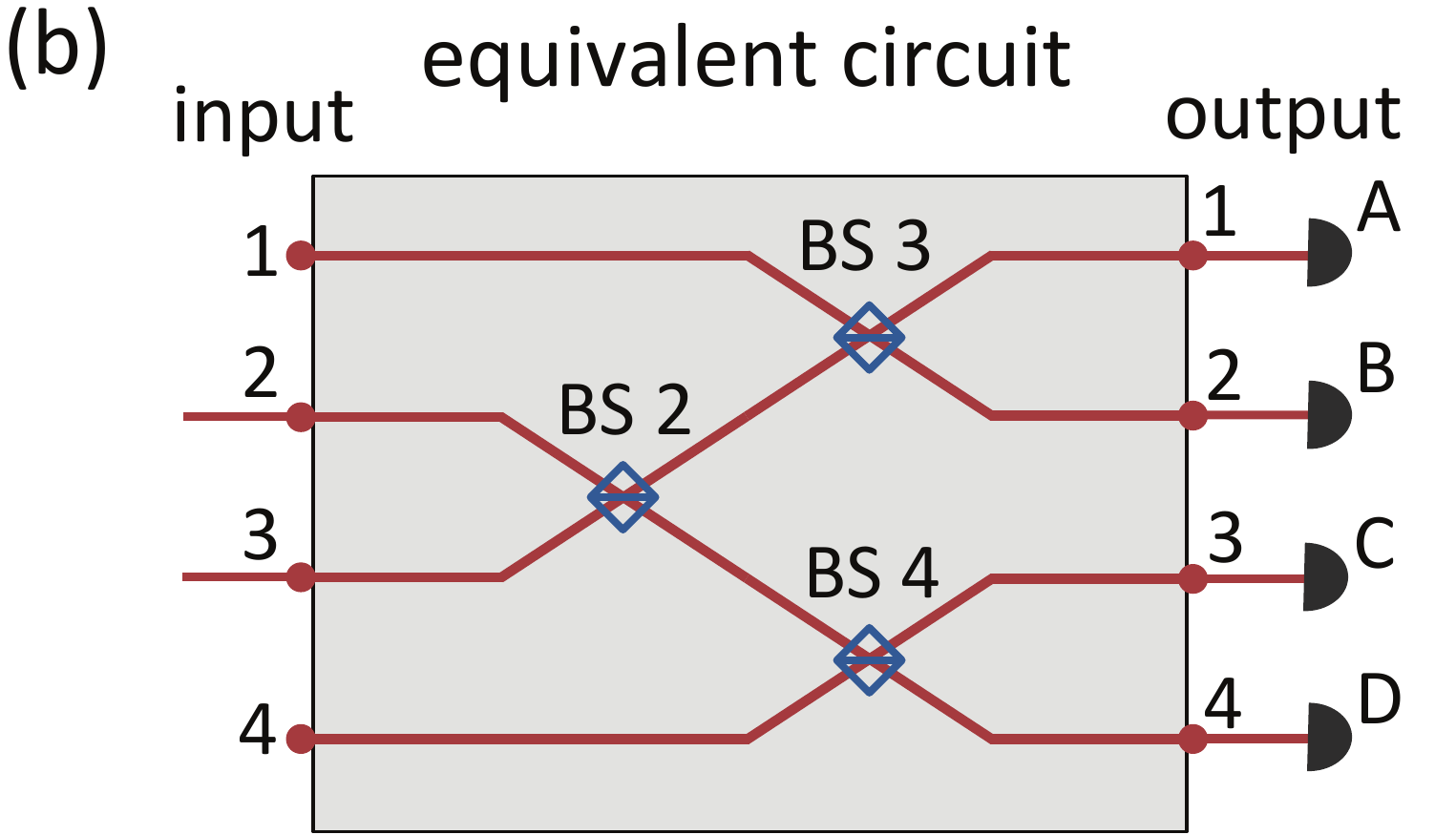}
		\end{minipage}%
		\quad
	%	\hspace{\dimexpr-\fboxrule-\fboxsep\relax}\fbox{%
		\begin{minipage}[c]{1\linewidth}
		\vspace*{2.0em}
	%	 \centering
			%{\footnotesize
			\vspace*{0.1em}
			\hspace*{1.5em} U=  
			\vspace*{0.5em}
	%		\centering
			$
			{\footnotesize
 %			\hspace*{1.1em}
			\left(
			\begin{matrix}
			ir_3 & ir_2t_3 &t_2t_3 &0\\
			t_3&-r_2r_3& ir_3t_2&0\\
			0 &ir_4t_2& -r_2r_4 &t_4\\
			0 &t_2t_4& ir_2t_4& ir_4
			\end{matrix}
			\right) 
			}  
			$
		\hfill
			\end{minipage}
%			}
	\end{minipage}
\vspace*{0.5em}
\caption{\label{fig1} (a) Measurement layout used to characterize the quantum light source. FBS: fiber beam splitter; PC: polarizer control; HWP: half-wave plate; BS: beam-splitter; detectors A-D. The measurement arrangement consists of an unbalanced interferometer beginning at FBS, containing a delay line (delay) and ending at BS2. A portion is fiber (black) and a portion is free space (red). The 4 detectors simulate two, two-photon number-resolving detectors. The HWP is used in a comparison experiment (see text). (b) Equivalent optical circuit without the initial FBS and the associated unitary matrix $U$; where $r_j$ and $t_j$ are reflection and transition coefficients for the three beamsplitters. \newline}
\label{schematic}
\end{figure}

A second-order correlation function can also be used to characterize the second-order interference of two input fields, and for non-entangled inputs determines their indistinguishability. 
Using an interferometer, replicas of the same field can be used, and when used
is often referred to as Hong, Ou, and Mandel (HOM) interferometry~\cite{hong_prl1987}, denoted $g^{(2)}_{\rm HOM}$. 
In our $g^{(2)}_{\rm HOM}$ measurement, unequal path lengths are used to match the arrival times of photons emitted at different times from a source onto a beam splitter, followed by two detectors to measure coincidences at the beam splitter outputs. 
If the photons are perfectly indistinguishable single photons, both exit the same beam-splitter port (BS2 in Fig.~\ref{schematic}a) making $g^{(2)}_{\rm HOM}[0] = 0$ ~\cite{hong_prl1987}. 
In general, $g^{(2)}_{\rm HOM}$ depends on the single-photon purity, and additional characterization is required for complete evaluation.
One option is to directly determine the single-photon purity through $g^{(2)}_{\rm HBT}$ ~\cite{Santori2002}.
Alternatively, a second measurement~\cite{flagg_prl2012} or series of measurements~\cite{kwiat1992} can be made in which the indistinguishablity is controllably varied; for example, by varying the polarization difference in each arm, and thus indirectly accessing the single photon purity~\cite{flagg_prl2012}. 
 
\section*{Simultaneous secord-order correlation measurements with number resolving detectors}
\vspace*{-1em}
In Fig.~\ref{schematic}a the Sg2 scheme is shown.
After the light source and polarizer controllers, it consists of an unbalanced interferometer, followed by two beam splitters (BS3, BS4) and four single-photon detectors. 
Each BS -- two-detector pair combination emulates a two-photon number-resolving photon detector. 

We operate under pulsed excitation, and use the notation $g^{(2)}[j]$ to denote the normalized integrated correlations between two detections $j$ pulses apart. 
In the Supplemental Material, we derive functions representing the single-photon purity $g^{(2)}_{\rm HBT}[0]$ and the indistinguishability $C$.
$C$ ranges from 0 for perfectly distinguishable photons to 1 for perfectly indistinguishable photons.
The results of these derivations are:

\begin{eqnarray}
g^{(2)}_{\rm HBT}[0]&=& \frac{g^{(2)}_{\rm auto}[0] + g^{(2)}_{\rm cross}[0] - \zeta(d) }{\zeta(0)} \nonumber \\
C & = & \frac{g^{(2)}_{\rm auto}[0] - g^{(2)}_{\rm cross}[0]}{\zeta(d)}, 
\label{autocorrelation}
\end{eqnarray}
where $d$ is the delay in the unbalanced interferometer. 
$\zeta$ accounts for temporal instabilities, in particular the source spectral jitter, and is discussed below.
$g^{(2)}_{\rm auto}[0]$ and $g^{(2)}_{\rm cross}[0]$ are auto- and cross-correlations of the two output fields.
While the Supplemental Material accounts for non-ideal beamsplitters in the interferometer, in Eqn.~\ref{autocorrelation} we assume the ideal case of 50:50 beamsplitters.
In previous non-number resolving two-detector schemes, just one function, $g^{(2)}_{\rm cross}$ is measured. 
Since we now have two functions, both the single photon purity and the indistinguishability can be simultaneously extracted from these quantities.
Because pairs of non-number resolving detectors are used, we simultaneously measure six of these second-order correlation functions: two are auto-correlations of the output field ($g^{(2)}_{\rm AB}$, $g^{(2)}_{\rm CD}$), and four are cross-correlations ($g^{(2)}_{\rm AC}$, $g^{(2)}_{\rm BC}$, $g^{(2)}_{\rm AD}$, $g^{(2)}_{\rm BD}$), where A, B, C and D denote the 4 detectors in Fig.~\ref{schematic}. 
We average them to form
\begin{eqnarray}
%\overbar{  }
g^{(2)}_{\rm auto}[j]  & = & \frac{1}{2} \left( g^{(2)}_{\rm AB}[j] + g^{(2)}_{\rm CD}[j]\right) \nonumber \\
g^{(2)}_{\rm cross}[j]  & = & \frac{1}{4}\left( g^{(2)}_{\rm AC}[j] + g^{(2)}_{\rm AD}[j]+g^{(2)}_{\rm BC}[j]+g^{(2)}_{\rm BD}[j]\right).
\label{SigmaDelta}
\end{eqnarray}

Discrete solid-state emitters can shift in energy with time, leading to spectral jitter~\cite{Kuhlmann2015,Thoma2016,Loredo2016}.
It is accounted for in Eqn.~\ref{autocorrelation} by the function $\zeta$, also discussed in the Supplemental Material. 
If present, this jitter degrades the indistinguishably, but will not degrade the single-photon purity. 
$\zeta$ is assumed to be a decaying exponential function of the form,
\begin{eqnarray}
\zeta(k) = 1 + \left( \zeta_0 -1 \right ) e^{-|k|/ \tau_1}
\label{zeta}
\end{eqnarray}
where $k$ refers to the number of pulses separating the generated photons, $\tau_1$ is the characteristic lifetime of the jitter here measured in pulse periods, and $\zeta_0$ is the value of $\zeta$ at zero delay.

\section*{Experimental realization of Sg2 correlation measurements}
\vspace*{-1em}
To demonstrate the Sg2 measurement we use photons emitted from a single InAs quantum dot (QD). 
Technical details about the QD device, how it is excited and how collection is made is in the Supplemental.
It is an emerging source of bright, single photon light~\cite{Gazzano2016}.
A single QD typically emits light with $g_{\rm HBT}^{(2)}[0]$ close to, but different from zero \cite{michler2000}.
QD photon indistinguishability can vary, but normalized values above 0.95 (within 5\% of perfectly indistinguishable) have been reported~\cite{gazzano2013,lu_natnanotech2013,wei2014,nowak2014}.

\begin{figure}[t!]
       \begin{minipage}[h!]{1\linewidth}
	\begin{center}
	\vspace*{1em}
	\includegraphics[width=4.5in]{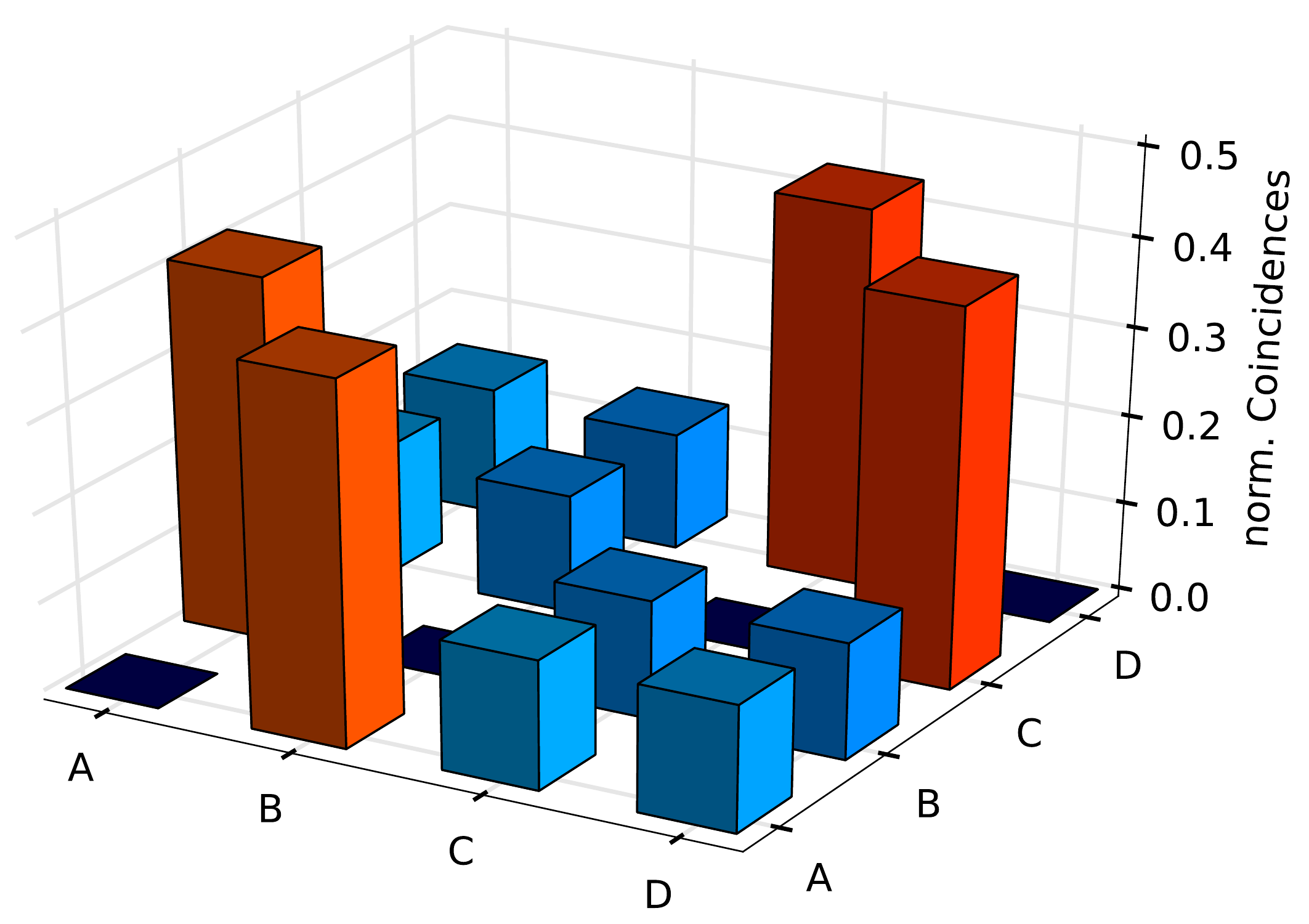}
	\end{center}
	\end{minipage}
\caption{Second-order characterization of quantum light; here from a single QD source. Normalized conditional detector counts are plotted for the four detectors at $j = \rm 0$.  Correlated detections on AD, AC, BC, BD (blue) represent cross-correlations. Correlations of the type AB and CD (red) are autocorrelations. \newline}
\label{4detector-g20}
\end{figure}

Using the set-up in Fig.~\ref{schematic}a, we measure the normalized second-order auto- and cross-correlations for each detector combination, as in Eqn.~\ref{SigmaDelta}. We normalize by the product of single-count probabilities $p_{l}p_{m}$, where $l$ and $m$ are the relevant detectors.
The result is shown in Fig. \ref{4detector-g20}. 

Correlations are grouped into two categories;  $g^{(2)}_{\rm cross}[0]$ (AC, AD, BC, BD) (blue, Fig.~\ref{4detector-g20} ) and $g^{(2)}_{\rm auto}[0]$ (AB and CD) (red, Fig.~\ref{4detector-g20}). 
One can qualitative observe the nonclassical properties of the source from the matrix in Fig.~\ref{4detector-g20}.
For instance, the larger values of the $g^{(2)}_{\rm auto}[0]$ terms (red) compared to $g^{(2)}_{\rm cross}[0]$ terms (blue) indicate that photons have a higher probability of leaving the same exit port of the interferometer beam-splitter. 
This {\it coalescence} indicates photon indistinguishability.

\begin{figure}[b!]
       %\begin{minipage}[h!]{1\linewidth}
	%\begin{center}
	%\includegraphics[width=3.1in]{Figures/Unknown-17.png}
	%\end{center}
	\vspace*{1em}
	%\end{minipage}
       \begin{minipage}[h!]{1\linewidth}
	 \begin{center}
	\includegraphics[width=4.5in]{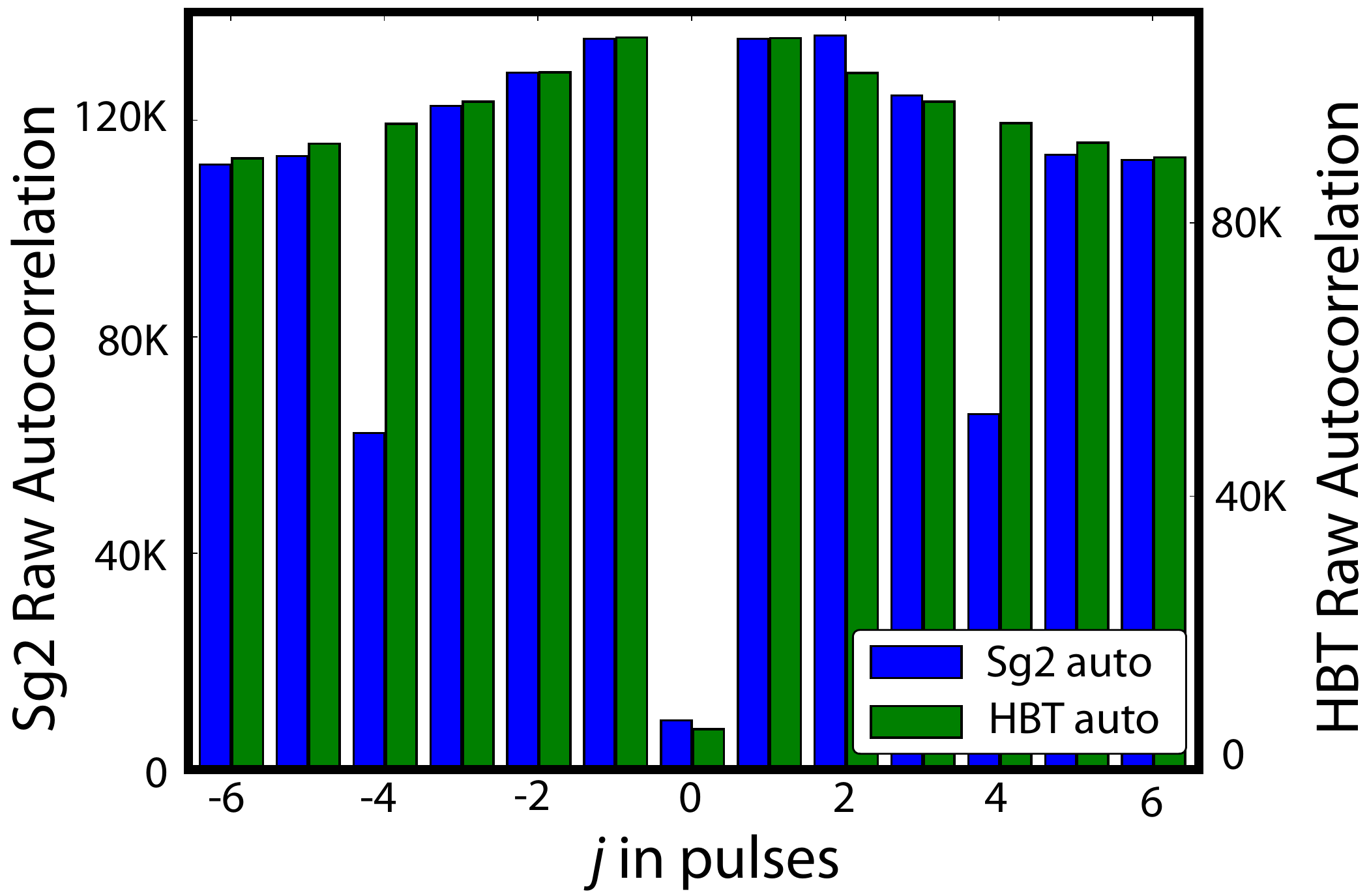}
	\end{center}
	\end{minipage}
%\vspace*{0.5em}
\caption{ A comparison of raw single-photon purity (autocorrelation) measurements, uncorrected for spectral jitter. The extracted autocorrelation from the four-detector Sg2 measurement (blue) compares well to the traditional HBT autocorrelation measurement (green). The two measurements are off-set laterally for clarity. The dip at $\rm \pm 4$ pulses is due to the interferometer (and single photon character of the source) (see text and supplemental for details). Using Eqn.~\ref{autocorrelation} in the text, $g^{(2)}_{\rm HBT}[0]$ can be found. \newline}
\label{HBT}
\end{figure}

We can quantitatively extract single photon purity and the indistinguishability from the data in Fig.~\ref{4detector-g20}. First, using Eqn.~\ref{autocorrelation} we can determine the single photon purity through $g^{(2)}_{\rm HBT}[0]$. 
The result is shown in blue in Fig.~\ref{HBT} where we plot the average of the unnormalized autocorrelations, uncorrected for the source jitter, {\it i.e.,} with $\zeta(k)$=1. 
This data captures the additional dynamics associated with jitter in the QD photon frequency on a longer time scale than the QD decay. 
Here, $g^{(2)}_{\rm HBT}[0] = 0.05(1)$.
$\zeta_0$=1.34 and $\tau_1$ = 3.64 laser pulses or 24.0~ns. 
The dips at $\rm \pm 4$ pulses are due to the 4 laser-pulse delay in the unbalanced interferometer (26.3~ns) and the single photon nature of the source. A more detailed description of this effect is found in the Supplemental.

Determining the photon indistinguishability follows in a straight-forward manner for the data in Fig.~\ref{4detector-g20} using Eqn.~\ref{autocorrelation}. We find $C = 0.61(1)$.  
Instead of directly determining $C$, in many situations it is more convenient to associate the fringe visibility $V$ with indistinguishability, particularly when a variable controlling indistinguishably is continuously varied; for instance, the polarization~\cite{kwiat1992}. For completeness, we determine it here using only one measurement set. We calculate $V$ using Eqn. S1 in the Supplemental Material, where it is determined directly from $g^{(2)}_{\rm HBT}[0]$ and $C$. We find  $V \rm = 0.58(1)$.

We compare these results with the traditional Hanbury Brown and Twiss (HBT) measurement\cite{HBT_nature1956}, albeit with 4 detectors instead of the usual two,
and the indistingishability results with the traditional measurement made in two steps. 
The traditional HBT measurement is done without the first beam splitter, {\it i.e.}, bypassing the interferometer and summing pair correlations over the 4 detectors. 
The comparison is shown in green in Fig.~\ref{HBT}.
%\textcolor{red}{\sout{ where we plot the unnormalized autocorrelations, uncorrected for the source jitter, {\it i.e.,} with $\zeta(k)$=1.}}
The measurements compare well. 
$g^{(2)}_{\rm HBT}[0] = 0.060(6)$ 
%\sout{($0.080(6)$ without the $\zeta$ correction)} 
using the standard HBT configuration.
Using the auto-correlation data and Eqn.~\ref{zeta}, we obtain the same values for $\zeta_0$ and $\tau_1$. 
The dip at 4 laser pulses is not present here because the unbalanced interferometer is not used.

To determine the indistinguishability in the traditional way, the value of $g^{(2)}_{\rm HBT}$ needs to be known and a second measurement with the interferometer determines $g^{(2)}_{\rm HOM}$. Here the indistinguishability is found from Ref.~\cite{flagg_prl2012} where the probability of coalescence is $C=1 + g^{(2)}_{\rm HBT}[0] - 2 g^{(2)}_{\rm HOM}[0]$.
The coalescence is $\rm 0.62(1)$, in good agreement with the new technique utilizing a single measurement.
 We also determine $V$ using the traditional method. It requires two measurements, where the half-wave plate (HWP) in Fig.~\ref{schematic} is set to rotate the polarization in one arm of the interferometer by $0$ or by $\pi/2$.   Using the traditional approach we find $V \rm = 0.58(9)$, equal to the Sg2 method result. The error is larger here because of the small data set for $g^{(2)}_{\rm HOM}$ when the HWP makes the two paths distinguishable.

\section*{Simultaneous full probability distribution measurement}
\vspace*{-1em}
To fully characterize the photon number distribution, higher order correlation measurements are generally required~\cite{goldschmidt_pra2013,Stevens2015}. Characterization to at least third order is necessary.
Using the standard HBT-type measurement with two single-photon detectors, second-order correlations determine photon number statistics to second order (N=2)~\cite{predojevic2014}. 
Using the Sg2 scheme, we can determine a probability distribution of the photon number states up to N=3.
In the QD light source used, a photon should be emitted from the light source every laser pump cycle (2 $\times$ 76 MHz); however, the system is only pumped to 70 \% of saturation. 
We determine the photon count rate at the fiber exiting our cryostat (just before the interferometer input) to be $3.08\cdot10^6$ cts/s, indicating the source efficiency (p$_{1}$) exiting the fiber is $0.029$.
p$_{1}$ is calculated based on the detector efficiencies and transmission through the interferometer, and these were measured to be 0.65~\%.
$p_0=1-p_1$ is then directly calculated to be $0.97$.
$p_2$ can be determined from $g^{(2)}_{\rm HBT}[0]$;  $g^{(2)}_{\rm HBT}[0] = 2p_{2}/p_{1}^{2}$ \cite{Stevens2015}, and $p_{2} = \rm 2.9 \cdot 10^{-5}$. 
The uncertainties on the above values are dominated by the long-term fluctuations in the set-up which we estimate to be 10 \%.
Zero 3 or 4 photon coincidences were measured for a trial number of $\rm 1.82 \cdot 10^{13}$, giving an upper limit of $p_3$ of $\rm 2.1\cdot10^{-6}$. 
The assumption of $p_0 \gg p_1 \gg p_2 \gg p_{N>2}$ holds, and the QD emission is described by a mixed state with a density matrix $\rm{||p||} = 0.97 p_{0,0} + 0.029 p_{1,1} + (2.9 \cdot 10^{-5}) p_{2,2} + \sum_{i>2} 0 p_{i,i} (+ 2.1 \cdot 10^{-6})$, where ${\rm p}_{n,m}=|f_n\rangle\langle f_m|$, and $n$, $m$ are elements of the density matrix. We note that the off-diagonal elements of p are expected to be zero ({\it i.e.,} no coherence between different number states).
%Similar measurements with two detectors~\cite{predojevic2014} or three detectors~\cite{stevens2014third} have been made.

\section*{Characterizing the efficiency}
\begin{figure}[h!]
 %      \begin{minipage}[l]{0.45\linewidth}
%       \hspace*{-.75em}
	\begin{center}
	\includegraphics[width=5.0in]{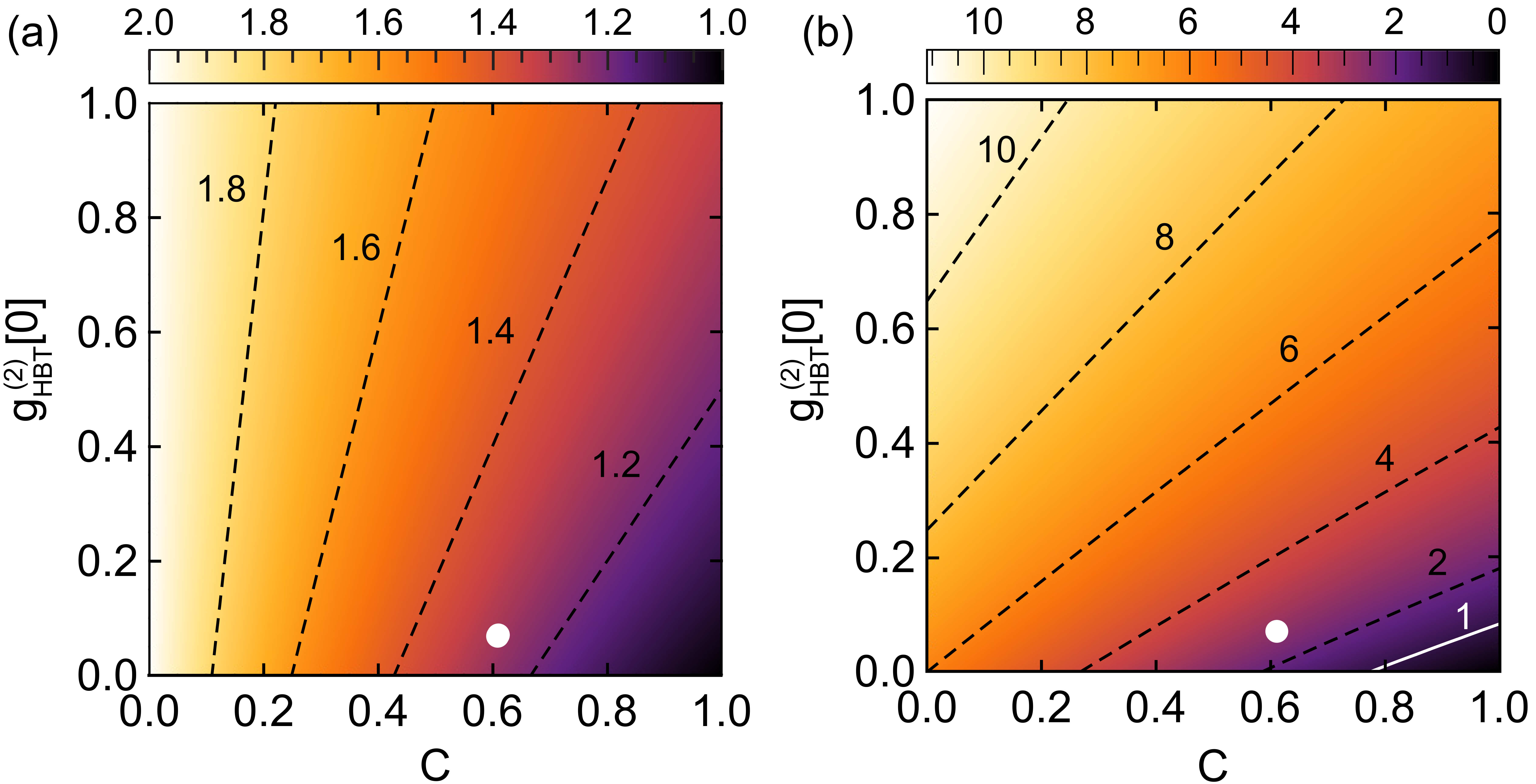}
	\end{center}
%	\end{minipage}
%	%	\hspace*{0.5em}
%	\quad
%       \begin{minipage}[r]{0.45\linewidth}
%        \hspace*{-.75em}
%%	 \begin{center}
%	\includegraphics[width=1.7in]{Figures/Fig-4b.png}
%%	\end{center}
%	\end{minipage}
%\vspace*{0.5em}
\caption{Efficiency of second-order correlation measurements.  (a) Scaling of the variance when measuring $C$. A photon-number resolved interferometer measurements reduces uncertainties faster than a non-photon number resolved method. The ratio of $\sigma^2 (C_{\rm trad})/\sigma^2 (C_{\rm Sg2})$ is shown (see text) as a function of $g^{(2)}_{\rm HBT}[0]$ and $C$, and here assumes that $g^{(2)}_{\rm HBT}[0]$ is known from a separate measurement. (b) Scaling of the scoring function $\frac{\sigma^2 \left ( g^{(2)}_{\rm trad}[0] \right ) + \sigma^2 \left ( C_{\rm trad} \right )}{\sigma^2 \left ( g^{(2)}_{\rm Sg2}[0] \right ) + \sigma^2 \left ( C_{\rm Sg2} \right )}$ when measuring $g^{(2)}_{\rm HBT}$ and $C$. Comparison between the Sg2 method and the method where two independent HBT and HOM measurements are used with two non-number resolving detectors~\cite{Santori2002}, expressed as ratios of uncertainty scoring functions, see text. Larger scale-bar numbers indicate a larger efficiency advantage of the Sg2 method. White dot: this source. \newline}  
\label{error}
\end{figure}

\vspace*{-1em}
In characterizing the efficiency of the Sg2 measurement, we first prove that the underlying photon-number resolving detection at the outputs of the unbalanced MZI provides more information about the input state than a conventional method. 
To do so, we compare the scaling of the variances, $\sigma^2$,  in measuring $C$ for each measurement ({\it i.e.}, per trial) of the photon-number resolved method and the conventional method. 
While we vary the indistinguishability and purity of the source, we assume that the purity is first known with certainty, since in the traditional approach it must be determined from a separate measurement.
Thus, the ratio of variances is 
\begin{equation}
\frac{\sigma^2 (C_{\mathrm{trad}}) }{\sigma^2 (C_{\mathrm{Sg2}})} = \frac{2 (g^{(2)}_{\mathrm{HBT}}+1)}{(g^{(2)}_{\mathrm{HBT}}[0]+C+1)},
\end{equation}
where subscripts {\it trad} and {\it Sg2} denote the two interferometer measurements with conventional or photon-number resolving detectors, respectively; see Fig.~\ref{error}a. 
$\sigma^2 ( C_{\mathrm{trad}})/\sigma^2 (C_{\mathrm{Sg2}})\geq1$ for all physically meaningful $g^{(2)}_{\mathrm{HBT}}$ and $C$, thus proving that the new method reduces the uncertainty in $C$ faster than the traditional method. 
The advantage is maximum at $g^{(2)}_{\mathrm{HBT}}[0]=1$, $C=0$, and reduces to unity ({\it i.e.} the two methods become identical) at $g^{(2)}_{\mathrm{HBT}}=0$, $C=1$. For the source investigated in this work, $\sigma^2 (C_{\mathrm{trad}})/\sigma^2 (C_{\mathrm{Sg2}})=1.28$, indicating that the photon-number resolving detection provides a 28~\% faster measurement than a traditional one. 

Next, we compare our Sg2 method to one of the popular two-step methods~\cite{somaschi2015,ding2015} for characterizing a single photon source, introduced by Santori, {\it et al.} \cite{Santori2002} which combines separate measurements of $g^{(2)}_{\rm HBT}$ and $C$. The efficiency is assessed by calculating the variances associated with $g^{(2)}_{\rm HBT}$ and $C$, and then using a scoring function of the form $\sigma^2 \left ( g^{(2)}[0] \right ) + \sigma^2 \left ( C \right )$.
The ratio of scoring functions is shown in Fig.~\ref{error}b, where it is assumed that in a Santori two-step method the total measurement time is evenly split between HBT and interferometer measurements. 
The Sg2 is more efficient most of the time except for those very near to perfect single-photon purity and indistinguishability.
For light with $g^{(2)}_{\mathrm{HBT}}[0]=0.5$, which is often used as a threshold for a discrete number-state source, the method is $\approx 7$ times faster when the indistinguishability is 0.5.
The maximum ratio is equal to 11.125, and the ratio for our source is equal to $\approx$~2.5.
This is because in a traditional method only a fraction of the total measurement can be used to determine single-photon purity, because the rest of the measurement should be used to determine indistinguishability. 
In addition, as established earlier, a photon-number resolved interferometer measurement reduces uncertainties faster than a traditional interferometer measurement.  
However, in the interferometer methods, the variance $\sigma^2 \left( g^{(2)}_{\mathrm{HBT}}[0] \right)$ scales proportionately to $g^{(2)}_{\mathrm{HBT}}[0]+ {\mathrm{const}}$, while in a traditional HBT method it scales with $g^{(2)}_{\mathrm{HBT}}[0]$. 
Therefore, an interferometer-based measurement of $g^{(2)}_{\mathrm{HBT}}[0]$ underperforms an HBT measurement for low $g^{(2)}_{\mathrm{HBT}}[0]$; see Fig.~\ref{error}b.
But of course, if the Sg2 method is supplemented with a separate HBT measurement, its efficiency would always be superior than a traditional HBT+interferometer measurement.

We would like to point out that other approaches exist.
For example, a visibility measurement can be made using two conventional MZI measurements in co- and cross-polarization configurations to measure $g^{(2)}_{\rm HBT}[0]$ and $C$.
Comparing the Sg2 method with this visibility method, the Sg2 approach is always more efficient.

Finally, since $n=2$ number resolving detectors provide a more efficient measurement method for second-order correlation measurements, perhaps higher order number resolving detectors would provide further improvements? This is not the case; no further efficiency benefit is provided by $n>2$ number resolving detectors.
Using $n>2$ number-resolving detectors would make higher order Glauber correlation measurements more efficient.
However, they would not improve the second-order correlation measurements, specifically $g^{(2)}_{\rm HBT}[0]$ and $C$.

\section*{Conclusions}
\vspace*{-1em}
%\paragraph{\normalfont\bfseries}{\bf Conclusions.} 
Using the light emission from a semiconductor QD structure as a test light source, we have demonstrated a new measurement approach allowing simultaneous measurement of the brightness, single photon purity and photon indistinguishability.
The approach uses an interferometer and 2 two-photon number-resolving detectors, here simulated by four detectors. 
The simultaneous second-order correlation (Sg2) approach proposed here eliminates any variation in source and experiment that may be present in independent measurements, and in nearly all cases does so with reduced uncertainty. 
The Sg2 measurement is especially efficient when the nonclassical light has less than ideal single-photon properties.
Finally, while $n=2$ number resolving detectors are the critical elements in this approach, higher-order number resolving detectors do not offer improved efficiency, although they would be useful to characterize higher order number states.

The Sg2 measurement scheme can be evaluated using a boson sampling approach, as seen in the equivalent circuit in Fig.~\ref{schematic}b. 
Boson sampling is the partial sampling of a bosonic circuit of an array of inputs and beam-splitters.
An important parameter in evaluating this circuit is the matrix permanent.
For simple systems like the one here, completely solving for the permanent of the matrix is trivial, but for large unitaries the solution of the permanent is difficult, in the $\rm \sharp P$ complexity class  \cite{valiant1979}, and boson sampling of the unitary can offer a tractable solution~\cite{spring2013,broome2013,ralph2013,tillmann2013,crespi2013}. The unitary matrix of this circuit, Fig.~\ref{schematic}c, can be used 
with the boson sampling model to determine the $g^{(2)}_{\rm HBT}[0]$ and $C$ values which most closely matches the Fock-state distribution~\cite{tillmann2013} found for the QD source. The $g^{(2)}_{\rm HBT}[0]$ and $C$ determined by the boson sampling model that best matches Fig.~\ref{4detector-g20} is within the uncertainty of those determined by the Sg2 approach. 
This Sg2 scheme could be incorporated into complex photonic circuits to assess the second-order correlation properties of the light.

Number-resolving detectors, as well as quantum-dot based single-photon sources like the one used here are emerging technologies that will likely have a strong impact in quantum measurement, experiments and systems. While we have shown that N=2 number resolving detectors will advance the characterization of quantum light, we believe they will also improve a diverse set of quantum optics experiments, for instance the boson sampling class of problems discussed above. We hope this work helps to further motivate such efforts.

This work was partially supported by the NSF PFC@JQI and the Army Research Laboratory.
We thank P. Senellart and T. Huber for helpful discussions and a critical reading of the manuscript.
S.V.P. thanks Pavel E. Samoylov for encouraging discussions.
E.A.G acknowledges support from the National Research Council Research Associateship program.
G.S.S. acknowledges support from Fulbright Austria - Austrian American Educational Commission through the Fulbright-University of Innsbruck Visiting Scholar program.

\bibliographystyle{apsrev4-1}
\bibliography{quantumlight}

\begin{thebibliography}{42}
\bibitem [{tes()}]{tes}%
  \BibitemOpen
  \href@noop {} {}\bibinfo {note} {For example, transition-edge sensors are
  currently available and have shown unsaturated photon-number resolution for
  low photon-number states (up to $\approx$~10 photons), but currently may lack
  temporal resolution for rapid sources such as the source discussed
  here.}\BibitemShut {Stop}%
\end{thebibliography}

\newpage


\section*{Supplementary material}

# Supplemental Material: Simultaneous, full characterization of a single-photon state

Tim Thomay,[1] Sergey V. Polyakov,[2] Olivier Gazzano,[1] Elizabeth Goldschmidt,[1] Vivien Loo,[1] and Glenn S. Solomon[1, 2, *]

[1]*Joint Quantum Institute, National Institute of Standards and Technology,*
*& University of Maryland, Gaithersburg, MD, USA.*
[2]*National Institute of Standards and Technology, Gaithersburg, MD, USA.*

## THE SEMICONDUCTOR LIGHT SOURCE

The nonclassical light source used is made from a semiconductor quantum dot (QD). The QD is made using the molecular-beam epitaxy crystal growth process. The QD portion of the sample is formed in MBE without lithography by the lattice-mismatch strain between InAs and the GaAs substrate. The QD density in the sample is approximately 10 QDs per $\mu m^2$. The QDs are located at the center anti-node of a 4-$\lambda$ planar distributed Bragg reflector (DBR) microcavity with 15.5 lower and 10 upper DBR pairs of GaAs and AlAs; the cavity mode is centered at a wavelength, $\lambda \approx 920$ nm. A schematic showing how the sample is pumped and light extracted is shown in Fig. S1. The QD emission used here is at 916.3 nm and the measured lifetime is 650 ps. A single-mode optical fiber is bonded to the

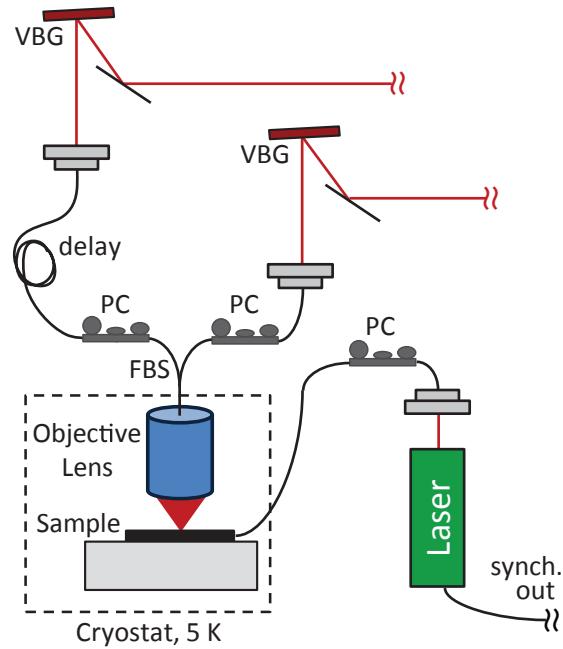

FIG. S1. The QD sample is in a 5K cryostat. It is excited through a cleaved edge by 895 nm laser light *via* a fiber glued to a cleaved sample edge. Light is collected from the top of the sample with lens that is fiber coupled. QD light in the fiber is coupled to an interferometer using a fiber beam splitter (FBS), with polarization controllers (PC), and is filter with volume Bragg gratings (VBG). The cryostat side of the interferometer is fiber (black), the section containing the VBG and beyond is free space (red).

cleaved [110] sample face to couple the excitation laser into the guided mode of the DBR cavity [S1]. Collection of light is made vertically from the top of the sample. The sample is maintained at approximately 5 K in a cryostat. Excitation *via* the side-bonded fiber is by a mode-locked Ti:sapphire laser with a repetition rate of 76.1 MHz and

---

* Corresponding author: gsolomon@umd.edu

approximately 8 ps pulse duration. However, the laser repetition rate is doubled so that the pulse spacing at the sample is 6.6 ns. This side excitation is made with 895 nm light that is absorbed into the roughened quantum well region intrinsic to the QD formation process, known as the wetting layer.

## SECOND-ORDER INTENSITY CORRELATIONS USING NUMBER-RESOLVING DETECTORS

To characterize the second-order correlation properties of the source with conventional (non-number resolving) detectors, two independent measurements are required. In one possible scheme, an unbalanced Mach Zehnder interferometer (MZI) is used with a polarizer in one arm. Measurements are recorded when the second-order interference is switched on and off by setting the polarization difference between the two arms to 0 and $\pi/2$, thus measuring the second-order interference visibility. The visibility can also be measured by a discrete, step-wise adjustment of the polarization [S2]. An alternative method again uses an unbalanced MZI in a maximal interference configuration, supplemented with an separate Hanbury-Brown Twiss (HBT) measurement, *i.e.*, a second-order correlation measurement with only one input to the beamsplitter.

Two important properties determined by these second-order correlation functions are the single photon purity, $g^{(2)}_{\text{HBT}}$ (the purity against multi-photon state content > 1) and the photon indistinguishability, $C$. $g^{(2)}_{\text{HBT}} = 0$ for a pure single photon source and $g^{(2)}_{\text{HBT}} = 1$ for a purely uncorrelated (Poissonian) source. $C = 1$ for fully indistinguishable photons and $C = 0$ for fully distinguishable ones. The physical meaning of $C = 1$ is that single photons will coalesce into a bi-photon state with unity probability when simultaneously arriving at two input arms of a beamsplitter. For the two conventional detectors with perfect 50/50 MZI beamsplitters, these properties are related by:

$$
\begin{aligned}
g^{(2)}_{\text{HOM},//}[0] &= \frac{1}{2}\left(1 + g^{(2)}_{\text{HBT}}[0] - C\right) \\
g^{(2)}_{\text{HOM},\perp}[0] &= \frac{1}{2} + \frac{1}{2}g^{(2)}_{\text{HBT}}[0] \\
V &= \left(g^{(2)}_{\text{HOM},\perp}[0] - g^{(2)}_{\text{HOM},//}[0]\right)/g^{(2)}_{\text{HOM},\perp}[0] \\
&= \frac{C}{1 + g^{(2)}_{\text{HBT}}[0]}.
\end{aligned}
\tag{S1}
$$

where $g^{(2)}_{\text{HOM},(//,\perp)}[0]$ in Eqn. S1 comes directly from the derivation of $C$ in Ref. [S3], and where $C = 1 + g^{(2)}_{\text{HBT}}[0] - 2g^{(2)}_{\text{HOM},//}[0]$. For no coalescence and in the weak photon intensity limit, $g^{(2)}_{\text{HOM},\perp} = g^{(2)}_{\text{HOM},//}$, and correctly, $C = 0$. If there is maximum coalescence, there is maximum interference in the MZI and a cross correlation, $g^{(2)}_{\text{HOM},//}[0]$, would give 0 except that the light source may not be a pure single photon source. If it is not a pure single photon source, a nonzero measure will occur but in the equation for $C$ it is accounted for by $g^{(2)}_{\text{HBT}}[0]$. $V$ in Eqn. S1 is the visibility. Determining indistinguishability from $V$ is the method traditionally used by many groups. However, as shown in Section V, it is the least efficient method to characterize a single-photon state due to uncertainty scaling and lost information.

We can make an autocorrelation measurement and a cross-correlation of the two fields, both taken at the output of the MZI. Because coalescence of indistinguishable fields at the beamsplitter produces bi-photon states, we can take advantage of number-resolving detectors in place of the conventional detectors to gain improved statistics. For the two output fields there are two autocorrelations ($g^{(2)}_{\text{auto}}[j]$), one for each number-resolving detector, and one cross-correlation ($g^{(2)}_{\text{cross}}[j]$), where $j$ is the time difference of detection events at the two detectors. Here, we use a pair of conventional detectors in a beamsplitter configuration to represent such a number-resolving detector (Fig. S2), and there are two $g^{(2)}_{\text{auto}}[j]$ terms and four $g^{(2)}_{\text{cross}}[j]$ terms. We find:

$$
\begin{aligned}
g^{(2)}_{\text{auto}}[0] &= \frac{1}{2}\left(1 + C + g^{(2)}_{\text{HBT}}[0]\right) \\
g^{(2)}_{\text{cross}}[0] &= \frac{1}{2}\left(1 - C + g^{(2)}_{\text{HBT}}[0]\right).
\end{aligned}
\tag{S2}
$$

$g^{(2)}_{\text{auto}}[0]$ in Eqn. S2 is the autocorrelation after the MZI and does not directly evaluate the single-photon purity of the source, while $g^{(2)}_{\text{HBT}}[0]$ is a traditional autocorrelation without the MZI. We note that these expressions are general for any detectors at the MZI output.

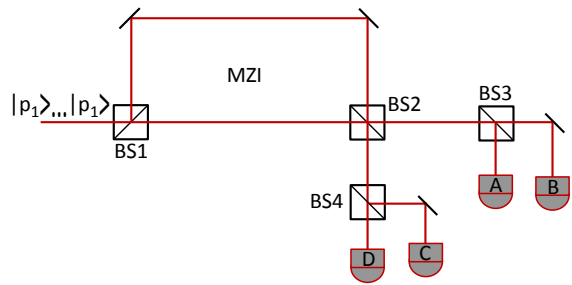

FIG. S2. An unbalanced Mach Zehnder interferometer (MZI) is used in a Hong-Ou-Mandel type set-up with output beamsplitter, BS2. Four detectors (A-D) are used instead of the usual two to simulate two N=2 number resolving detectors.

To calculate $g^{(2)}_{\text{HBT}}(0)$ and $C$ simultaneously for two input photon wave packets, we assume that the two-photon probability, $p_2$ in each wave packet is much smaller than the single-photon probability, $p_1$ in the wave packets and that the zero-photon probability, $p_0 \approx 1$, i.e., $p_2 \ll p_1 \ll 1$. Here, we use a pair of conventional detectors in a beamsplitter configuration to represent such a number-resolving detector, and there are two $g^{(2)}_{\text{auto}}[j]$ terms and four $g^{(2)}_{\text{cross}}[j]$ terms, as shown in Fig. S2.

a. **Number-resolving second-order correlations including source bunching.** A blinking-like effect has been observed in the literature [S4, S5] and observed in our results here. It is likely the result of the quantum dot (QD) exciton energy shifting in energy (jitter). We therefore assume that an *inherent* $g^{(2)}[0]$ is present but altered by an overall bunching effect around $j = 0$. We identify the inherent $g^{(2)}[0]$ by $g^{(2)}_{\text{HBT}}[0]$, and it is defined in the usual way,

$$p_2^{(i)} = \frac{1}{2} g^{(2)}_{\text{HBT}}[0] \left(p_1^{(i)}\right)^2 \tag{S3}$$

for all shots $i$. $p_2^{(i)}$ is the two-photon probability at the input.

We also consider a time-difference dependent second-order correlation, $\zeta(k)$ to model the QD energy shift in and out of the filter window of the general form of a covariance,

$$\zeta(k) = \frac{\left\langle p_1^{(i)} p_1^{(i+k)} \right\rangle}{\langle p_1 \rangle^2} \tag{S4}$$

where $(k)$ is the time difference between photons emitted by a source. $\zeta(k \to \infty) = 1$ since at large time differences there are no correlations. As defined, $\zeta(k) = \zeta(-k)$.

The beamsplitters transmission and reflectance in intensity are $T$ and $R$. In the case where one photon is incident on each of the two input ports of the second MZI beamsplitter, the probability of photons simultaneous exiting the same output port of the final MZI beamsplitter (BS2) is [S6]

$$t_{B2} = \left[1 - \left(R_2^2 + T_2^2 - 2T_2 R_2\right)\right] C + 2R_2 T_2 \left(1 - C\right) \tag{S5}$$

$t_{B2}$ is the full output transmission when two photons exit the same outport. For only one port, such as the one going to detectors A and B this should be halved. Note further that if $T_2 = R_2 = \frac{1}{2}$ then $t_{B2} = \frac{1}{2}(1+C)$.

Using Eqns. S3, S4, and S5 we can form various second-order coincidence probabilities to determine second-order auto and cross correlations. For example:

$$p_{\text{AB}}[0] = 2\langle p_1 \rangle^2 \left( \frac{g^{(2)}_{\text{HBT}}[0]}{2} \zeta(0) \left(T_1^2 T_2^2 + R_1^2 R_2^2\right) + \zeta(d) T_1 R_1 \frac{t_{B2}}{2} \right) T_3 R_3$$

$$p_{\text{AC}}[0] = \langle p_1 \rangle^2 \left( \frac{g^{(2)}_{\text{HBT}}[0]}{2} \zeta(0) \left(T_1^2 + R_1^2\right) 2 T_2 R_2 + \zeta(d) T_1 R_1 (1 - t_{B2}) \right) R_3 R_4$$

$$p_{\text{AB}}[\pm d] = \langle p_1 \rangle^2 \left( \frac{g^{(2)}_{\text{HBT}}[0]}{2} \zeta(0) 2 T_1 R_1 T_2 R_2 + \zeta(d) \left(T_1^2 T_2^2 + R_1^2 R_2^2\right) + \zeta(2d) T_1 R_1 T_2 R_2 \right) T_3 R_3$$

$$p_{\text{AC}}[+d] = \langle p_1 \rangle^2 \left( \frac{g^{(2)}_{\text{HBT}}[0]}{2} \zeta(0) 2 T_1 R_1 T_2^2 + \zeta(d) \left(T_1^2 + R_1^2\right) + \zeta(2d) T_1 R_1 R_2^2 \right) R_3 R_4 \quad \text{(S6)}$$

where $d$ is the delay in the MZI. For the general case of $j \neq 0, \pm d$,

$$p_{\text{AB}}[j] = \langle p_1 \rangle^2 \left( \zeta(j) \left(T_1^2 T_2^2 + R_1^2 R_2^2\right) + \zeta(j+d) T_1 R_1 T_2 R_2 + \zeta(j-d) T_1 R_1 T_2 R_2 \right) T_3 R_3$$

$$p_{\text{AC}}[j > d] = \langle p_1 \rangle^2 \left( \zeta(j) \left(T_1^2 + R_1^2\right) T_2 R_2 + \zeta(j+d) T_1 R_1 R_2^2 + \zeta(j-d) T_1 R_1 T_2^2 \right) R_3 R_4$$

$$p_{\text{AC}}[0 < j < d] = \langle p_1 \rangle^2 \left( \zeta(j) \left(T_1^2 + R_1^2\right) T_2 R_2 + \zeta(j+d) T_1 R_1 R_2^2 + \zeta(j-d) T_1 R_1 R_2^2 \right) R_3 R_4 \quad \text{(S7)}$$

We now normalize the $p_2$'s to get $g^{(2)}[j]$. Noting that

$$p_A = \langle p_1 \rangle (T_1 T_2 + R_1 R_2) R_3$$
$$p_B = \langle p_1 \rangle (T_1 T_2 + R_1 R_2) T_3$$
$$p_C = \langle p_1 \rangle (T_1 R_2 + R_1 T_2) R_4$$
$$p_D = \langle p_1 \rangle (T_1 R_2 + R_1 T_2) T_4$$

Using detectors A and B as an example, $g^{(2)}_{\text{AB}}[j]$ is

$$g^{(2)}_{\text{AB}}[j] = \frac{p_{\text{AB}}[j]}{\langle p_A \rangle \langle p_B \rangle} = \frac{p_{\text{AB}}[j]}{\langle p_1 \rangle^2 [T_1 T_2 + R_1 R_2]^2 T_3 R_3}. \quad \text{(S8)}$$

While this is a rigorous definition for $g^{(2)}_{\text{AB}}[j]$, we note that $p_{\text{AB}}$ is more typically normalized by $p_{\text{AB}}[j+l]$ for large $l$'s for convenience. However, this practice leads to systematic errors when dynamics are present as the use of $p_{\text{AB}}[j+l]$ assumes $p_{\text{AB}}$ is time independent.

Now the normalized $g^{(2)}_{d_1 d_2}[j]$ can be found for detectors $d_1, d_2 \in$ A, B, C, D. For 50:50 lossless beamsplitters, the normalizations are $1/\langle p_A \rangle \langle p_B \rangle = 16/\langle p_1 \rangle^2$, and Eqns. S6 reduced to:

$$g^{(2)}_{\text{AB}}[0] = \frac{1}{2} \left[ g^{(2)}_{\text{HBT}} \zeta(0) + \zeta(d)(1 + C) \right]$$

$$g^{(2)}_{\text{AC}}[0] = \frac{1}{2} \left[ g^{(2)}_{\text{HBT}} \zeta(0) + \zeta(d)(1 - C) \right]$$

$$g^{(2)}_{\text{AB}}[d] = g^{(2)}_{\text{AC}}[d] = \frac{1}{4} \left[ g^{(2)}_{\text{HBT}} \zeta(0) + 2\zeta(d) + \zeta(2d) \right]$$

$$g^{(2)}_{\text{AB}}[j > d] = g^{(2)}_{\text{AC}}[j > d] = \frac{1}{4} \left[ 2\zeta(j) + \zeta(j+d) + \zeta(j-d) \right] \quad \text{(S9)}$$

For the four detector configuration we obtain two auto correlation terms and four cross correlation terms:

$$g^{(2)}_{\text{auto}}[j] = \frac{1}{2} \left( g^{(2)}_{\text{AB}}[j] + g^{(2)}_{\text{CD}}[j] \right)$$

$$g^{(2)}_{\text{cross}}[j] = \frac{1}{4} \left( g^{(2)}_{\text{AC}}[j] + g^{(2)}_{\text{AD}}[j] + g^{(2)}_{\text{BC}}[j] + g^{(2)}_{\text{BD}}[j] \right). \quad \text{(S10)}$$

Using only the $j = 0$ case in Eqn. S6,

$$g^{(2)}_{\text{HBT}}[0] = \frac{1}{\zeta(0)} \left( g^{(2)}_{\text{auto}}[0] + g^{(2)}_{\text{cross}}[0] - \zeta(d) \right)$$
$$C = \frac{1}{\zeta(d)} \left( g^{(2)}_{\text{auto}}[0] - g^{(2)}_{\text{cross}}[0] \right) \tag{S11}$$

where $g^{(2)}_{d_1 d_2}[0]$ is the experimentally measured value and includes the dynamics, while $g^{(2)}_{\text{HBT}}[0]$ is again the intrinsic value. There is additional information on $g^{(2)}_{\text{HBT}}[0]$ with the $j = \pm d$ terms that can be found from $g^{(2)}_{\text{auto}}[d] + g^{(2)}_{\text{cross}}[d]$:

$$g^{(2)}_{\text{HBT}}[0] = \frac{1}{\zeta(0)} \left[ 2 \left( g^{(2)}_{\text{auto}}[d] + g^{(2)}_{\text{cross}}[d] \right) - 2\zeta(d) - \zeta(2d) \right] \tag{S12}$$

Notice here that by subtracting substantially similar values, the uncertainty increases compared to $g^{(2)}_{\text{HBT}}[0]$ evaluations.

The function $\zeta$ models any jitter in the system. With no jitter $\zeta(k) = 1$ for all $k$. When jitter is present, there is a systematic error in evaluating an intrinsic $g^{(2)}[0]$. If random jitter is present $\zeta$ would naturally be a decaying exponential of the form,

$$\zeta(k) = 1 + (\zeta_0 - 1) e^{-|k|/\tau_1} \tag{S13}$$

where $\tau_1$, measured in pulse periods, is the characteristic lifetime of the QD photon energy jitter, and $\zeta_0$ is the value of $\zeta$ at delay zero. The spectral diffusion is related to the photon emission rate, hence $k$ is the time difference between photon emission. At $k = \pm d$ the statistics are altered and seen in Eqn. S9.

To account for the experimentally observed jitter, we incorporate S13 into the correlations represented in S9. A particularly useful correlation is for detection differences $|j| > d$ on detectors $d_1$ and $d_2$,

$$g^{(2)}_{d_1 d_2}[j > d] = \frac{1}{4} \left[ 2\zeta(j) + \zeta(j+d) + \zeta(j-d) \right]$$
$$= 1 + \frac{1}{4} (\zeta_0 - 1) e^{j/\tau_1} \left( 2 + e^{d/\tau_1} + e^{-d/\tau_1} \right)$$
$$= 1 + \frac{1}{2} (\zeta_0 - 1) e^{-|j|/\tau_1} (1 + \cosh(d/\tau_1)) \tag{S14}$$

For the specific case above, the functional form does not depend on whether $d_1 d_2$'s are auto or cross correlations. Once $\zeta(j)$ is recovered, $C$ and $g^{(2)}_{\text{HBT}}[0]$ can be determined from Eqn. S11.

## MEASURING AUTOCORRELATIONS AT THE TIME DELAY OF THE INTERFEROMETER

In Fig. 3 the dip in the data using the interferometer at $\pm 4$ laser pulses is due to a combination of the unbalanced interferometer delay and the single photon character of the light source. There are four possible path combinations through the interferometer to achieve specific "start"-"stop" detector time differences, $j$. When $j$ equals the interferometer path difference (here at $j = \pm 4$ pulses) and the light source is a pure single photon source, one of the possible paths does not occur. This can be seen from Eqn. S9, line 3, where the pre-factor is reduced to $1/4$. Since $g^{(2)}_{\text{HBT}}$ is related to the sum of both $g^{(2)}_{\text{auto}}$ and $g^{(2)}_{\text{cross}}$, in the $j = \pm 4$ pulse bins in Fig. 3 the $g^{(2)}_{\text{HBT}}$ should drop by a factor of 2 for a pure single-photon source. Since the source is not purely antibunched the reduction will be slightly smaller. Such a condition could be used to determine the single-photon purity but there will be large errors in this measurement because of subtraction of two relatively large values to obtain a smaller one.



\end{document}